\documentclass[aps,prl,twocolumn,floatfix,superscriptaddress]{revtex4}

\usepackage{graphicx}
\usepackage{dcolumn}
\usepackage{bm}

\newcommand{\diffl}[2]{\frac{d #1}{d #2}}
\newcommand{\dfrac}[2]{\displaystyle\frac{#1}{#2}}

\newcommand{\kB}{k_{\text{B}}}
\newcommand{\thetaD}{\theta_{\text{D}}}

\newcommand{\fTQ}{f_{\text{TQ}}}

\begin{document}


\title{Equation of state and strength of diamond in high pressure ramp loading}

\date{December 21, 2019, revisions to August 31, 2021 
   -- LLNL-JRNL-800622}

\author{Damian~C.~Swift}
\affiliation{%
   Lawrence Livermore National Laboratory,
   7000 East Avenue, Livermore, California 94551, USA
}
\author{Olivier~Heuz\'e}
\affiliation{%
   CEA/DAM-\^Ile de France,
   Bruy\`eres-le-Ch\^atel, F-91297 Arpajon Cedex, France
}
\author{Amy~Lazicki}
\author{Sebastien~Hamel}
\author{Lorin~X.~Benedict}
\author{Raymond~F.~Smith}
\author{James~M.~McNaney}
\affiliation{%
   Lawrence Livermore National Laboratory,
   7000 East Avenue, Livermore, California 94551, USA
}
\author{Graeme~J.~Ackland}
\affiliation{%
   Department of Physics, University of Edinburgh,
   Edinburgh, EH9~3JZ, Scotland, UK
}

\begin{abstract}
Diamond is used extensively as a component in high energy density experiments,
but existing equation of state (EOS) models do not capture its observed response
to dynamic loading.
In particular, in contrast with first principles theoretical EOS models, 
no solid-solid phase changes have been detected, and no general-purpose EOS
models match the measured ambient isotherm.
We have performed density functional theory (DFT) calculations of the diamond phase
to $\sim$10\,TPa, well beyond its predicted range of thermodynamic stability, 
and used these results as the basis of a Mie-Gr\"uneisen EOS.
We also performed DFT calculations of the elastic moduli, and calibrated
an algebraic elasticity model for use in simulations.
We then estimated the flow stress of diamond by comparison with the
stress-density relation measured experimentally in ramp-loading experiments.
The resulting constitutive model allows us to place a constraint on the
Taylor-Quinney factor (the fraction of plastic work converted to heat)
from the observation that diamond does not melt on ramp compression.
\end{abstract}

\keywords{equation of state, electronic structure}

\maketitle

\section{Introduction}
Carbon is of course a key element in astrophysics and for terrestrial life,
and has been the subject of inumerable experimental and theoretical studies.
Its properties are particularly exotic because of the different forms of
interatomic bonding it exhibits.
Electronic structure studies have predicted multiple solid phases
to occur under compression, as discussed further below.
Although graphite is the equilibrium structure at STP, the diamond phase
is essentially stable for most purposes.
Both of these structures are covalently bonded, but diamond has a 
remarkably high shear modulus and strength;
its flow behavior is not understood.

Diamond components are used widely in high energy density (HED) experiments.
It is an efficient choice as the fuel capsule and ablator for
inertial confinement fusion (ICF) targets,
and is a common choice for the tamper and ablator for high pressure
studies of material properties, such as {\it in situ} x-ray diffraction
experiments \cite{Rygg2020}.
Diamond is also the material of choice for advanced static compression
systems intended to reach at least 1\,TPa \cite{Dubrovinsky2012,Dewaele2018,Jenei2018}.
The plastic flow behavior at high pressures is important in the design of these devices.
Extensive studies have been performed of its high pressure properties,
including measurements of its response to ramp loading to $\sim$5\,TPa
\cite{Bradley2009,Smith2014},
and x-ray diffraction to $\sim$2\,TPa \cite{Lazicki2021}.

Theoretical studies of the phase diagram of C have a long history
\cite{oldcalcs}.
Calculations of the graphite phase are relatively challenging 
using density functional theory (DFT) 
\cite{Hohenberg64,Kohn65,Perdew92,White94}
because of the difficulty in 
capturing the van der Waals forces between adjacent graphene sheets,
and most EOS models take diamond as the ambient phase.
Using electronic structure calculations resembling current techniques,
there has been a series of predictions of pressure-induced
transitions out of the diamond structure.
Diamond was predicted to transform to the R8 structure around 500\,GPa
\cite{Clark1995}.
More recently, in a multiphase EOS also including the liquid phase,
diamond was predicted to transform to BC8 around 1\,TPa \cite{Correa2008}.
Quantum molecular dynamics (QMD) simulations were performed to $\sim 1.4$\,TPa,
and compared with shock measurements at Sandia National Laboratories' Z facility
\cite{Knudson2008};
these results were used in the construction of a general-purpose EOS model,
{\sc sesame} 7834 \cite{ses7834}.
A particularly thorough EOS model has been constructed comprising four solid phases
and the liquid \cite{Martinez2012,Benedict2014}, again with a transition to BC8 around
1\,TPa.
For the elasticity,
electronic structure calculations have been used to calibrate an algebraic
model for the variation of the shear modulus of polycrystal diamond to
1.7\,TPa \cite{Orlikowski2007},
and the variation of the elastic moduli with pressure and temperature 
to 500\,GPa \cite{Valdez2012}.

Unfortunately, no current EOS and strength model captures the observed
behavior adequately.
The clearest indication is that no phase changes have been observed or
inferred in samples of carbon (initially diamond) up to 5\,TPa.
We have found no general purpose EOS model that reproduces measurements
of the ambient isotherm \cite{Occelli2003}.

As C has not been observed to change from the diamond structure
under ramp compression to at least 5\,TPa, we perform electronic structure
calculations and construct an EOS model for this structure to higher pressures,
for use in simulations of experiments in this regime.
Previous electronic structure calculations for EOS and elasticity 
have used a variety of 
numerical prescriptions and computer programs (Table~\ref{tab:prescrips}).
In the present work, we use a prescription consistent with those we
employed for our previous studies, 
e.g. \cite{Swift_Si_2001,Swift_NiTi_2005,Swift_NiAl_2007,Swift_RuRh_2019}.
DFT has systematic errors that often make it inappropriate to use directly as an EOS.
Our approach is to use DFT to obtain algebraic functions that can be adjusted to
improve the accuracy, and then use these functions to construct 
a model of the EOS and elastic moduli.

\section{Electronic structure calculations}
Energies and stresses used to construct the EOS and elastic moduli were
obtained from calculations of the ground state of the electrons with
respect to fixed ions,
for a series of different values of the lattice parameters.
These calculations were performed using 
non-local pseudopotentials to represent the inner electrons on each atom,
and a plane wave expansion to represent the outer electrons,
solving the Kohn-Sham DFT equations \cite{Hohenberg64,Kohn65,Perdew92,White94}
with respect to the Schr\"odinger Hamiltonian.
\footnote{%
For high-pressure EOS studies, we have generally found that 
the local density approximation (LDA) to the exchange-correlation functional and 
higher-order functionals such as the generalized gradient approximation (GGA)
produce results of similar accuracy, typically with a similar discrepancy in predicting
the mass density at STP, but of opposite sign.
LDA requires less computational effort, so we used it in this work.}
Pulay corrections to the ground state energy and stress were included,
although they were small.
Exchange-correlation interactions with the pseudopotential core were ignored;
as was found previously \cite{Benedict2014},
the cold curve at high compressions was found to asymptote to the
all-electron atom-in-jellium result, suggesting that these core corrections
are not significant.
The stress tensor on the lattice cell was obtained 
from the ground state wavefunctions using the Hellmann-Feynman theorem.

The pseudopotentials used were generated by the Troullier-Martins method
\cite{Troullier91} with the $K$-shell electrons
treated as core and the outer four treated explicitly as valence.
The wavefunctions were evaluated at $10^3$ regularly-spaced points in
reciprocal space, reduced by the symmetry of the crystal lattice 
\cite{Monkhorst76}.
The eight-atom diamond cubic lattice cell was used.
A plane-wave cutoff of 900\,eV was sufficient to converge the ground states
to $\sim$1\,meV/atom or better.

\begin{table}
\caption{Prescriptions used for electronic structure calculations.
   XC is the exchange-correlation functional, Kohn-Sham local density
   approximation (LDA) or generalized gradient approximation (GGA).
   $E_c$ is the plane wave cutoff energy.}
\label{tab:prescrips}
\begin{tabular}{|l|l|l|r|l|}\hline
 & {\bf XC} & {\bf $k$-points} & $E_c$ & {\bf software} \\
 & & & (eV) & \\ \hline
Clark et al \cite{Clark1995} & GGA & 4 ($3\times 3\times 3$) & 408 & {\sc cetep} \\
Orlikowski et al \cite{Orlikowski2007} & GGA & $20\times 20\times 20$ & 950 & {\sc abinit} \\
Correa et al \cite{Correa2008} & GGA & $10\times 10\times 10$ & 950 & {\sc abinit} \\
Knudson et al \cite{Knudson2008} & GGA & $\sim 3\times 3\times 3$ & 500 & {\sc vasp} \\
Benedict et al \cite{Benedict2014} & GGA & $20\times 20\times 20$ & 1300 & {\sc vasp} \\
Valdez et al \cite{Valdez2012} & LDA & $4\times 4\times 4$ & 680 & {\sc quantum} \\
 & & & & {\sc espresso} \\
This work & LDA & $10\times 10\times 10$ & 900 & {\sc castep}  \\
\hline\end{tabular}
\end{table}

\section{Equation of state}
The EOS model was constructed from a cold curve fitted to the electronic
structure calculations for the isotropically-compressed diamond cell,
with ion-thermal energy obtained from a Debye model \cite{Debye1912} in which the variation
of the Debye temperature $\thetaD$ with compression was calculated from
the Gr\"uneisen parameter obtained from the cold curve and its derivatives.
The EOS model is intended primarily for applications around the principal isentrope,
so the electron-thermal energy and the high-temperature reduction in 
ionic heat capacity to $3\kB/2$ were not included.
This approach avoids having to calculate the phonon density of states
as we have for other materials \cite{Swift_Si_2001,Swift_NiAl_2007},
and which is computationally much more intensive.
This simplification is justified as, in the solid, 
the ion-thermal contribution is a relatively small correction to the 
dominant contribution of the cold curve to the EOS,
and in the present work we are not concerned with 
phase boundaries where small corrections may matter.

We used the Burakovsky-Preston form \cite{Burakovsky2004} 
of the relationship between the Gr\"unseisen parameter $\Gamma(\rho)$
and the cold curve,
\begin{equation}
\Gamma(\rho)=\dfrac{\frac{B'(\rho)}2-\frac 16-\frac t2\left[1-\frac{p(\rho)}{3B(\rho)}\right]}{1-\frac {2t}3\frac{p(\rho)}{B(\rho)}}
\end{equation}
where $p$ is the pressure, $B$ the bulk modulus, and $B'$ its pressure
derivative.
The ion-thermal EOS has been found to be represented most accurately by
a value of $t$ which increases from 0 to 2 with compression
\cite{Burakovsky2004}.
We constructed EOS models with $t=1$ and $t=2$, and found that the latter
gave better agreement with the principal shock Hugoniot,
and so used the Vashchenko-Zubarev relation
\cite{Vashchenko1963} (corresponding to $t=2$) for subsequent work.
This choice is reasonable since the ion-thermal contribution 
to the principal shock Hugoniot is smaller at lower pressure.
For comparison, we also show some results for an EOS model constructed with
$t=1$, corresponding to the Dugdale-MacDonald form \cite{Dugdale1953}.

The Gr\"uneisen parameter is the logarithmic derivative of the Debye temperature,
\begin{equation}
\Gamma(\rho)=\dfrac{\rho}{\thetaD}\diffl{\thetaD}\rho,
\end{equation}
and so the Debye temperature can be expressed as
\begin{equation}
\thetaD(\rho)=\thetaD(\rho_r)G(\rho)
\label{eq:coldthetad}
\end{equation}
where
\begin{equation}
G(\rho)\equiv\exp\int_{\rho_r}^\rho \dfrac{\Gamma(\rho')}{\rho'}d\rho'
\end{equation}
and $\rho_r$ is some reference density.

In order to perform the further differentiation necessary to calculate
$B$ and $B'$, the cold curve was fitted with analytic functions.
Several functions were tried, and a modified version of the 
Vinet form \cite{Vinet1987}
was found to fit the cold curve over the widest range:
\begin{equation}
e(\rho)=e_c\phi[a(\rho)]+e_0 : \phi(a)=-e^{-a}\left(1+a+\frac 1{20}a^3\right)
\end{equation}
where 
\begin{equation}
a(\rho)=\dfrac{\left(\rho_0/\rho\right)^{\frac 13}-1}{l},\quad
l(\rho)=P_n[\eta(\rho)] :
\eta(\rho)=\dfrac \rho{\rho_0}-1
\end{equation}
and $P_n$ is a polynomial, here quadratic.
The fitting parameters (Table~\ref{tab:vinetparams})
are the coefficients of $P_n$, $\rho_0$, $e_c$, and $e_0$.
The modifications remove the need for material-specific
constants such as the atomic weight, used to calculate the
Wigner-Seitz radius, and thus give a more convenient relation 
for describing arbitrary $e(\rho)$ data, given that the scaling parameter
$l$ is fitted to reproduce the data in any case.
By expressing the cold curve in terms of macroscopic quantities such as mass density,
rather than atomic-level quantities such as Wigner-Seitz radius, it also becomes
simpler to apply the Vinet model to compounds and mixtures, rather than just elements.
In the original Vinet form, $l$ is a constant;
we found it necessary to generalize it to a low-order polynomial
in order to fit the cold curve of diamond C to multi-terapascal pressures.

\begin{table}
\caption{Modified Vinet parameters fitted to electronic structure calculations.}
\label{tab:vinetparams}
\begin{tabular}{|r|r|r|r|r|r|}\hline
 $\rho_0$ & $e_c$ & $l_0$ & $l_1$ & $l_2$ & $e_0$ \\
 (g/cm$^3$) & (MJ/kg) & & ($\times 10^3$) & ($\times 10^5$) & (MJ/kg) \\ \hline
3.64504 & 99.7657 & 0.29836 & -1.41913 & -5.42538 & -1149.17 \\
\hline\end{tabular}
\end{table}

In the present work, we took $\thetaD$ inferred at STP, 
and Eq.~\ref{eq:coldthetad} was used to calculate $\thetaD(\rho)$.
With the EOS model constructed in this way, the pressure for the observed STP
mass density and temperature was 13.8\,GPa.
It is desirable to apply a correction in order to bring the STP state into
as close agreement as possible with observation.
In constructing other EOS, we followed previous work \cite{Akbarzadeh1993} that argued 
that, since the origin of the discrepancy is the inaccuracy of the DFT
calculation, it makes sense to correct the latter rather than, for example,
scaling the mass density, as has been done for other EOS.
Accordingly, we have previously applied a pressure correction,
implented as an energy correction linear in specific volume.
The problem with this approach is that, in DFT, the pressure correctly
asymptotes toward zero as mass density does the same; this limit is violated
by the simple pressure correction.
Instead, we hypothesize a correction of the form 
\begin{equation}
\Delta e=\alpha\rho^\beta
\quad\Rightarrow\quad
\Delta p=\alpha\beta\rho^{\beta+1}.
\end{equation}
We took $\beta=1$, giving $\alpha=-p_{ai}(\rho_r,T_r)/\rho_r^2$
to bring the pressure to the desired STP value.

\subsection{High pressure performance}
Key thermodynamic loci were constructed from the EOS model,
for comparison with experimental data and with other models.
We compare with wide-ranging semi-empirical EOS models,
and also with the recent, multiphase EOS model \cite{Benedict2014} described above.
The semi-empirical models were constructed using 
empirical forms for the cold compression curve and ion-thermal energy,
adjusted to reproduce high pressure data as available at the time
of construction,
blending into Thomas-Fermi theory \cite{tf} at high temperature and compression.
The STP state in the multiphase model had a mass density 3.216\,g/cm$^2$,
which is significantly lower than the 3.51\,g/cm$^3$ observed in diamond.
For a more meaningful comparison, this model was adjusted in the same way
as described above to give a more accurate STP state;
it is referred to below as `Benedict, adjusted.'
The empirical EOS were {\sc sesame} models 7830 \cite{ses7830} and 7834 \cite{ses7834}.

\begin{figure}
\begin{center}\includegraphics[scale=0.72]{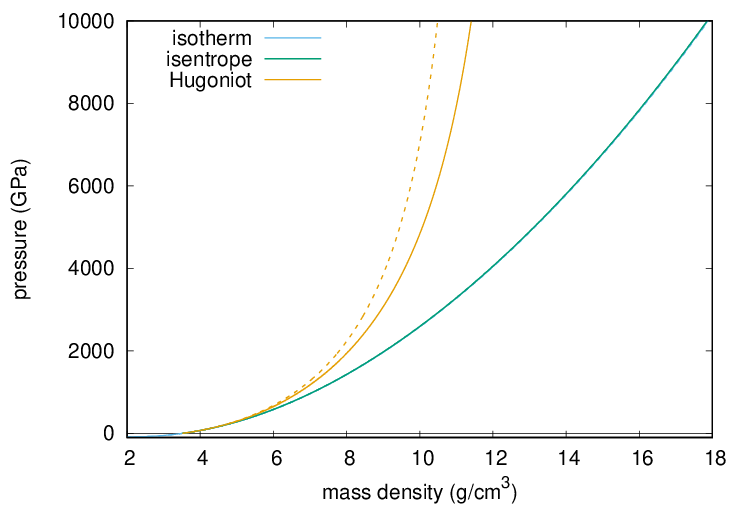}\end{center}
\caption{State loci from EOS models using Vashchenko-Zubarev (solid) and 
   Dugdale-MacDonald (dashed) $\Gamma(\rho)$.
   The isotherm and isentrope are almost identical on this scale.}
\label{fig:locidp}
\end{figure}

\begin{figure}
\begin{center}\includegraphics[scale=0.72]{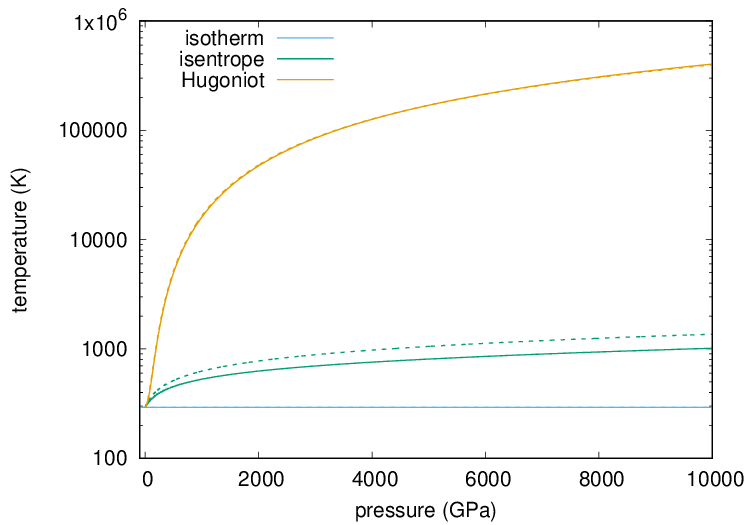}\end{center}
\caption{State loci from EOS models using Vashchenko-Zubarev (solid) and 
   Dugdale-MacDonald (dashed) $\Gamma(\rho)$.}
\label{fig:locipt}
\end{figure}

\begin{figure}
\begin{center}\includegraphics[scale=0.72]{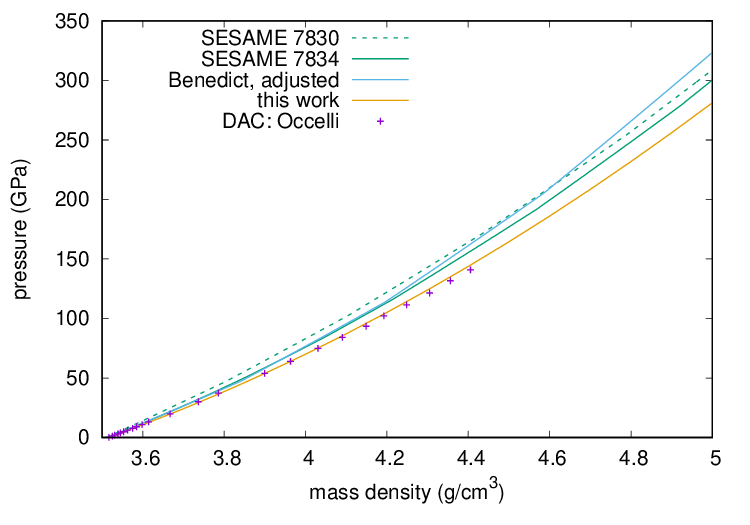}\end{center}
\caption{Comparison of ambient isotherm from different EOS models
   \cite{ses7834,ses7830,Benedict2014}
   with DAC measurements \cite{Occelli2003}.}
\label{fig:isocmp}
\end{figure}

\begin{figure}
\begin{center}\includegraphics[scale=0.72]{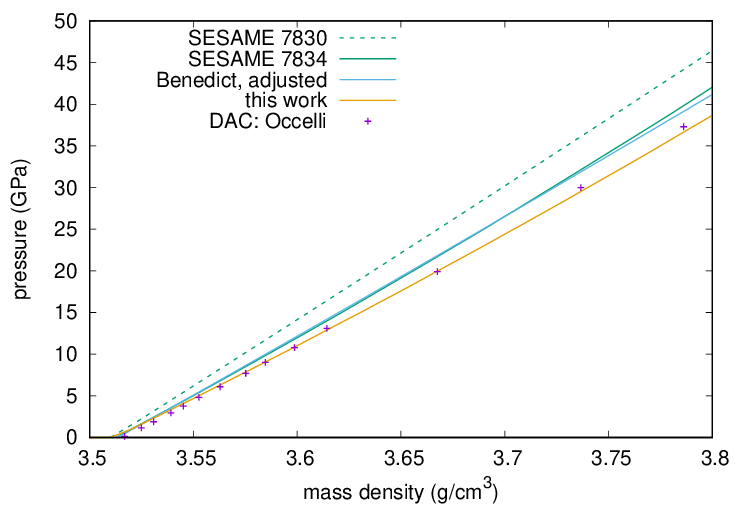}\end{center}
\caption{Comparison of ambient isotherm from different EOS models
   \cite{ses7834,ses7830,Benedict2014}
   with DAC measurements \cite{Occelli2003}
   (detail at low pressure).}
\label{fig:isocmp1}
\end{figure}

In pressure-density space, the choice of Gr\"uneisen model affected only
the principal Hugoniot;
in pressure-temperature space it affected only the principal isentrope
(Figs~\ref{fig:locidp} and \ref{fig:locipt}).
The ambient isotherm lies close to diamond anvil cell (DAC) measurements
\cite{Occelli2003}, closer than other reported EOS models
(Figs~\ref{fig:isocmp} and \ref{fig:isocmp1}).
The agreement may be even better than shown,
as it has been argued \cite{Dorogokupets2007} that the
pressure deduced in the DAC data should be raised 10\,GPa between $\sim$50 and 140\,GPa.

\begin{figure}
\begin{center}\includegraphics[scale=0.72]{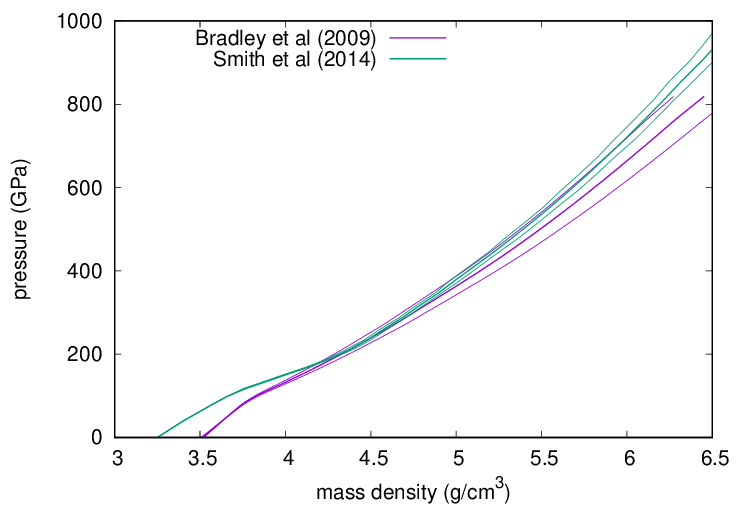}\end{center}
\caption{Comparison of ramp-loading data from samples of different
   initial density, to different peak pressures \cite{Bradley2009,Smith2014}.
   Thin lines are the loci of $1\sigma$ uncertainty for density reported in the original publications,
   which dominates the reported pressure uncertainty.}
\label{fig:isencmp0}
\end{figure}

Ramp-loading experiments have been reported on samples of full density to $\sim$800\,GPa
at the {\sc omega} laser facility \cite{Bradley2009},
and on deposited samples of reduced density to $\sim$5\,TPa at 
the National Ignition Facility (NIF) \cite{Smith2014}.
In the absence of material strength, ramp loading should sample states along
the principal isentrope of the sample;
strength acts to increase the normal stress at a given mass density.
The nominal stress-density relation deduced for the reduced-density material
lies above that from the full-density material everywhere except around 250\,GPa,
where they coincided.
This observation could be interpreted as a higher thermal pressure 
from the lower-density samples, or, unexpectedly, a higher strength.
However, taking the experimental uncertainties into account, the stress-density relations
are consistent with each other once the lower-density samples have been compacted to
full density (Fig.~\ref{fig:isencmp0}).

\begin{figure}
\begin{center}\includegraphics[scale=0.72]{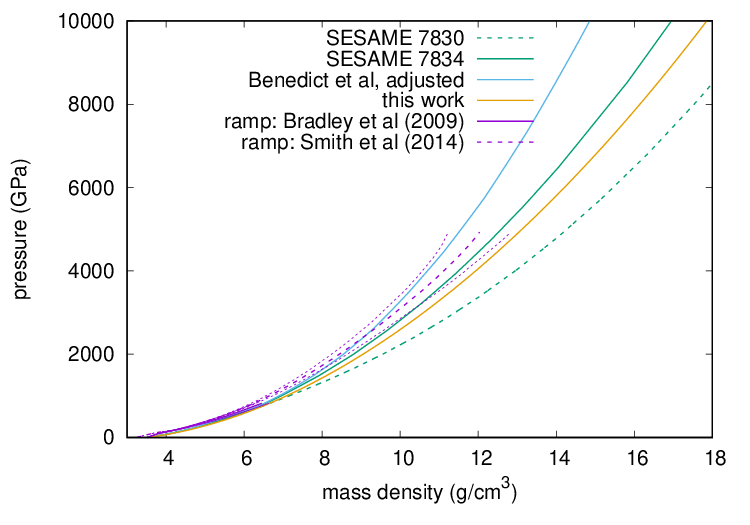}\end{center}
\caption{Comparison of principal isentrope from different EOS models
   \cite{ses7834,ses7830,Benedict2014}
   with ramp-loading data \cite{Bradley2009,Smith2014}.
   Thin lines are the loci of $1\sigma$ uncertainty in the ramp data.}
\label{fig:isencmp}
\end{figure}

\begin{figure}
\begin{center}\includegraphics[scale=0.72]{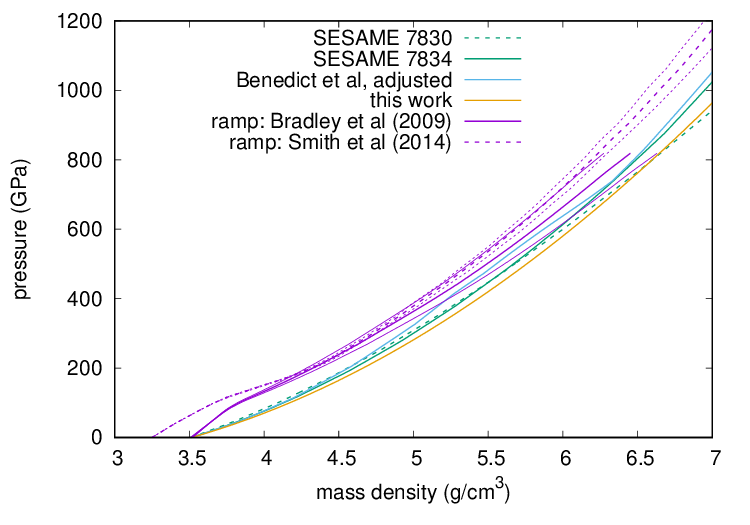}\end{center}
\caption{Comparison of principal isentrope from different EOS models
   \cite{ses7834,ses7830,Benedict2014}
   with ramp-loading data \cite{Bradley2009,Smith2014}
   Thin lines are the loci of $1\sigma$ uncertainty in the ramp data.
   (detail at low pressure).}
\label{fig:isencmp1}
\end{figure}

Before we consider the effect of material strength,
the only assessment we can make is whether the isentrope lies below the
ramp data, plotted as normal stress as a function of mass density,
by an amount plausible to account for by the strength.
The present EOS model has an isentrope lying in between those of the two
empirical models, consistent with (i.e. lying below) the ramp-loading data,
and implying a higher strength than would be consistent with
{\sc sesame} 7834.
The isentrope from the five-phase model passes through the nominal NIF ramp
data around 2.4\,TPa, although it always lies below the upper uncertainty bound of
the data, implying that the strength would have to decrease significantly
as the pressure increased beyond 1\,TPa.
If the nominal NIF data are correct, this would imply that the five-phase model
was inconsistent, as the strength would have to be negative, which is unphysical.
(Figs~\ref{fig:isencmp} and \ref{fig:isencmp1}).

\begin{figure}
\begin{center}\includegraphics[scale=0.72]{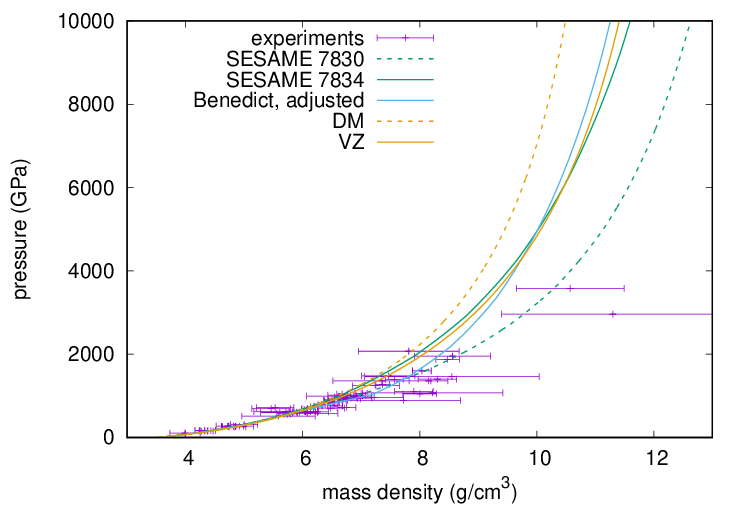}\end{center}
\caption{Comparison of principal shock Hugoniot from different EOS models
   \cite{ses7834,ses7830,Benedict2014} and constructions using
   Vashchenko-Zubarev (VZ) or Dugdale-MacDonald (DM) $\Gamma(\rho)$
   with shock data
   \cite{Knudson2008,Pavlovskii1971,Kondo1983,Bradley2004,Nagao2006,Brygoo2007,Hicks2008,McWilliams2010}.}
\label{fig:hugdp1}
\end{figure}

\begin{figure}
\begin{center}\includegraphics[scale=0.72]{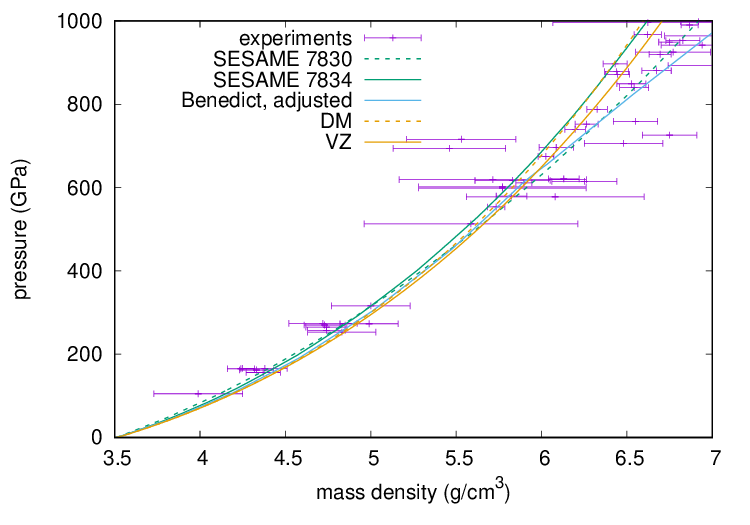}\end{center}
\caption{Comparison of principal shock Hugoniot from different EOS models
   \cite{ses7834,ses7830,Benedict2014} and constructions using
   Vashchenko-Zubarev (VZ) or Dugdale-MacDonald (DM) $\Gamma(\rho)$
   with shock data 
   \cite{Knudson2008,Pavlovskii1971,Kondo1983,Bradley2004,Nagao2006,Brygoo2007,Hicks2008,McWilliams2010}
   (detail at low pressure).}
\label{fig:hugdp2}
\end{figure}

\begin{figure}
\begin{center}\includegraphics[scale=0.72]{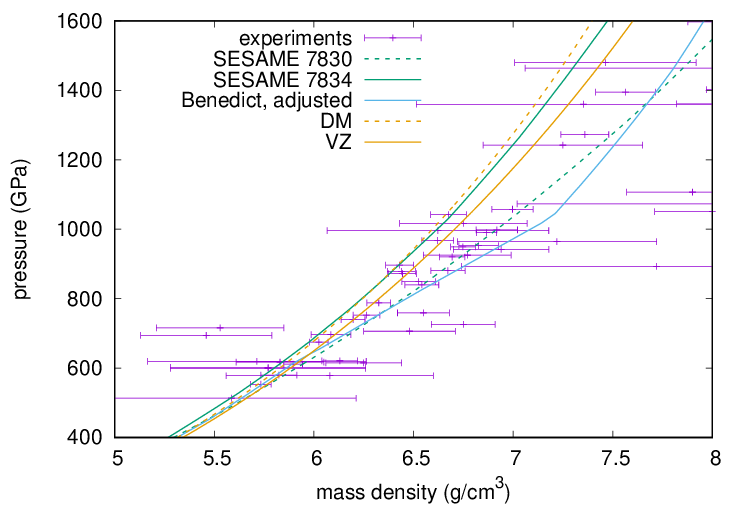}\end{center}
\caption{Comparison of principal shock Hugoniot from different EOS models
   \cite{ses7834,ses7830,Benedict2014} and constructions using
   Vashchenko-Zubarev (VZ) or Dugdale-MacDonald (DM) $\Gamma(\rho)$
   with shock data
   \cite{Knudson2008,Pavlovskii1971,Kondo1983,Bradley2004,Nagao2006,Brygoo2007,Hicks2008,McWilliams2010}
   (detail around shock melting).}
\label{fig:hugdp3}
\end{figure}

The present EOS model is intended primarily for ramp loading,
and did not include a melt curve, high temperature ion-thermal effects,
or electron-thermal energy, and thus would not be expected to reproduce
the shock Hugoniot at pressures much over 1\,TPa.
It is nevertheless useful to evaluate its performance at lower pressures,
and interesting to check its behavior at higher pressures, where for
example a ramp load could steepen into a shock in a less than optimal
experiment, as many are.
Shock Hugoniot data exhibit more random scatter than ramp measurements,
and experimental data up to the onset of shock melting were reasonably
consistent with any of the EOS models considered.
At higher pressures, the models constructed with Gr\"uneisen parameters
$t=0$ or 1 were much stiffer than any other EOS models, which is why we
eliminated them from further use.
Of the previous EOS models considered, only the five-phase model included an
explicit melt transition, clearly evident along the principal Hugoniot
between $\sim$800 and 1050\,GPa.
The Hugoniot from
the EOS model constructed with the Vashchenko-Zubarev Gr\"uneisen relation,
i.e. $t=2$, 
lies between those from the more recent empirical model {\sc sesame 7834}
and the five-phase model, even passing almost exactly where these Hugoniots
cross at around 5\,TPa to lie between them at even higher pressures.
This surprising consistency suggests that the new EOS model should be
at least as reasonable a choice when modeling shock formation from a
ramp load as either of the other EOS models.
It does not reproduce shock data in detail around melting though it seems to be 
as good a choice as any other EOS for shock states at higher pressures;
more work would be needed to develop an adequate model of the melt transition
out of the diamond phase where it may be metastable.
(Figs~\ref{fig:hugdp1} to \ref{fig:hugdp3}.)

\section{Constitutive model}
In theoretical and computational terms, it is far more feasible to
predict the EOS and elastic moduli than the plastic flow behavior,
because the length and time scales involved in plasticity are much less
tractable to {\it ab initio} computation, and the flow mechanisms are 
poorly understood.
Previous work on diamond for dynamic loading applied a time-independent
model developed for plastic flow in metals \cite{Orlikowski2007}.
It is commonly assumed, but not rigorously tested, that the flow stress
scales with the shear modulus.
Here we use electronic structure calculations to predict the variation
with compression of the elastic moduli, 
and hence the texture-averaged shear modulus.
We then consider observations of the normal stress to estimate the
variation of flow stress with compression.

In empirical studies of material strength, it is common to express the
elastic moduli and flow stress as functions of the pressure and temperature.
Electronic structure calculations are more amenable to predicting 
material properties as a function of mass density rather than pressure.
At constant pressure, increasing the temperature usually causes expansion;
expansion usually reduces the bulk and shear moduli,
and so the effect of temperature becomes conflated with that of compression.
Usually, expressing the shear modulus in terms of mass density reduces
the variation with temperature, sometimes to the level where it can be
ignored.
For this reason, 
we constructed the strength model in terms of mass density,
but we also provide a calibration in terms of pressure for use in 
software implementations which do not allow a compression-based model.

\subsection{Elastic moduli}
The variation of elastic moduli with compression was predicted by
calculating the ground state stress tensor as the diamond lattice cell was
distorted from its equilibrium shape.
Uniaxial compression along the $[100]$ direction gives $c_{11}$ and $c_{12}$,
and shearing in any $\{100\}$ plane gives $c_{44}$.
The stress tensor was calculated for several strains from each isotropic
configuration: a uniaxial strain of $\pm 0.1, 0.2, 0.5$, and a 
shear strain of $\pm 0.1, 0.2$.
The diamond lattice was calculated to remain stable with respect to
shear induced by uniaxial deformation, i.e. $c_{11} > c_{12}$, for
$\rho<20$\,g/cm$^3$, i.e. $p<12$\,TPa.
The calculated variation of $c_{44}$ was reasonably smooth and monotonic
up to 9.5\,g/cm$^3$, or 2.2\,TPa, at which point the calculated values became
scattered.
Ignoring widely-scattered and unconverged points, the variation
of elastic moduli with compression $\mu\equiv\rho/\rho_0-1$ 
was fitted by
\begin{equation}
\dfrac{c_{11}-c_{12}}2=b_0+b_1\mu+b_2\mu^2
\end{equation}
where $b_0=470\pm 4$\,GPa, $b_1=475\pm 6$\,GPa, and $b_2=-123\pm 1.4$\,GPa;
and
\begin{equation}
c_{44}=a_0+a_1\mu^\alpha
\end{equation}
where $a_0=521\pm 64$\,GPa, $a_1=1834\pm 88$\,GPa, and $\alpha=1.30\pm 0.09$.

The main purpose of this work is to improve models of diamond for use
in simulations of HED experiments, performed using multi-physics
radiation hydrocodes which typically include a small set of 
isotropic strength models, i.e. with a scalar shear modulus $G$ and
flow stress $Y$.
It is thus a practical advantage if the constitutive behavior of diamond
can be represented using an existing such model, even if it is not the
most natural or general representation of the behavior.
For uniaxial loading, the full, tensorial constitutive model can always
be represented by an effective isotropic model of sufficient complexity.
The Steinberg-Guinan model \cite{SteinGuin} is widely used in
dynamic loading simulations.
However, as normally formulated, the underlying prescription for the 
variation of $G$ with pressure and compression is represented with
a single parameter, and is known to be inaccurate at high pressures.
Instead, we consider a modification to the Steinberg-Guinan model
for large compressions \cite{KlepeisBales}, 
which has been used previously for diamond \cite{Orlikowski2007}.
This model has been implemented with some slight but important
differences in different hydrocodes, and for clarity we define below
the precise form of the model used in this work:
\begin{eqnarray}
G &=& G_0\left\{f(\rho)G_l+\left[1-f(\rho)\right]G_h-B(T-T_0)\right\} \\
f(\rho) &=& \left\{1 + \exp\left[\alpha (\eta(\rho) - \eta_c)\right]\right\}^{-1} \\
G_l &=& 1+A_l p \left[\eta(\rho)\right]^{-1/3} \\
G_h &=& A_h + M_h \eta(\rho) \\
\eta(\rho) &=& \rho/\rho_0
\end{eqnarray}
where $G_l$ is the usual Steinberg-Guinan term for pressure-hardening
and the term in $B$ similarly for thermal softening,
$G_h$ is the high pressure modification with parameters $A_h$ and $M_h$,
and $f$ provides a smooth transition between the two with parameter $\eta_c$
representing the compression at the transition.

Another important subtlety is that, for samples of different microstructural
texture, the effective isotropic strength model differs.
The previous diamond strength model \cite{Orlikowski2007} was constructed
for polycrystalline material assuming an isotropic texture.
Diffraction experiments on polycrystalline samples 
typically use single-crystal diamonds as structural components to avoid
superimposing diffraction rings from the diamond on the signal from the sample.
Measurements of loading waves transmitted through diamond of different
microstructure, $O(0.1)$\,mm thick, have indicated that plasticity
is simpler for $\{110\}$ crystals than for other orientations,
in the sense that less structure is observed around the elastic-plastic
transition in surface velocimetry \cite{Eggert2012}.
More detailed studies of single-crystal plasticity in diamond would be needed 
to understand this behavior and and reproduce it in simulations.
For this reason, we have generally preferred to use $\{110\}$ crystals for
our experiments at NIF and {\sc omega}.

\begin{figure}
\begin{center}\includegraphics[scale=0.72]{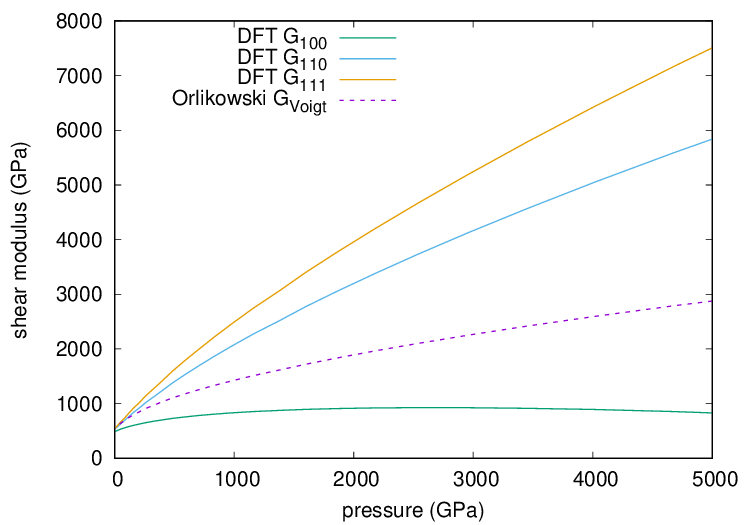}\end{center}
\caption{Predicted effective shear modulus as a function of pressure
   for different crystal orientations, compared with previous polycrystalline
   average \cite{Orlikowski2007} commonly used in simulations of HED experiments.}
\label{fig:pgxxxcmp}
\end{figure}

\begin{figure}
\begin{center}\includegraphics[scale=0.72]{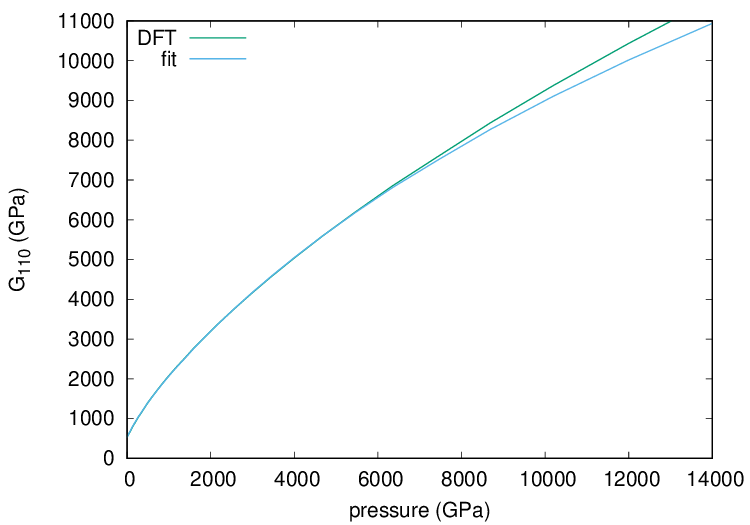}\end{center}
\caption{Improved Steinberg-Guinan fit to effective shear modulus for 
   $\{110\}$ crystals.}
\label{fig:g110fit}
\end{figure}

The effective shear modulus for $\{110\}$ crystals is
$G_{110} = (G_{100} + 3 G_{111})/4$,
where $G_{100}=(c_{11}-c_{12})/2$
and $G_{111}=c_{44}$
(Fig.~\ref{fig:pgxxxcmp}).
We fitted the parameters in the improved Steinberg-Guinan model above
to $G_{110}$ as calculated from the elastic moduli (Table~\ref{tab:isgparams}).
The parameters of the transition function $f$ 
provide such a broad transition that $G$ is sensitive to the
high pressure parameters throughout the low pressure region:
the parameter $G_0 < G(p=0)$.
The model reproduced the DFT calculations of $G_{110}$ almost exactly
to 6\,TPa, with deviations increasing to a few percent by $\sim$10\,TPa.
(Fig.~\ref{fig:g110fit}.)

\begin{table}
\caption{Parameters for the improved Steinberg-Guinan strength model.
   The parameters from Orlikowski \cite{Orlikowski2007} are the fit to DFT for comparison,
   rather than the values adjusted to match experimental observations of elastic moduli
   at low pressures.}
\label{tab:isgparams}
\begin{center}
\begin{tabular}{|c|r|r|r|}\hline
{\bf parameter} & $\{110\}$ & \multicolumn{2}{c|}{\bf polycrystalline} \\ \cline{3-4}
 & & Hill average & Orlikowski \cite{Orlikowski2007} \\ \hline
$G_0$ (GPa) & 386.5 & 552.5 & 529.0 \\
$A_l$ (1/GPa) & $0.212 \times 10^{-2}$ & $0.443 \times 10^{-2}$ & $0.43\times 10^{-2}$ \\
$A_h$ & -3.75 & -0.520 & -0.596 \\
$M_h$ & 5.41 & 1.368 & 1.588 \\
$\alpha$ & 0.67 & 0.574 & 3.47 \\
$\eta_c$ & 1.64 & 0.420 & 0.88 \\
\hline\end{tabular}
\end{center}
\end{table}

\begin{figure}
\begin{center}\includegraphics[scale=0.72]{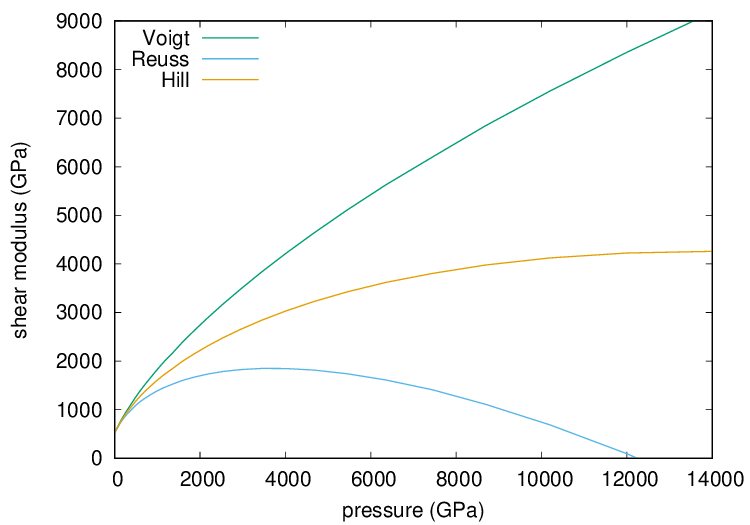}\end{center}
\caption{Predicted shear modulus for equiaxed polycrystalline diamond
   with alternative averaging methods.}
\label{fig:gpolycmp}
\end{figure}

For comparison with the previous strength model \cite{Orlikowski2007},
we also calculated the polycrystalline average shear modulus,
by the limiting approximations of Reuss (constant stress) and Voigt
(constant strain), and also Hill (average of the two).
These different approximations
diverge by tens of percent above a few hundred gigapascals,
indicating that texture and its evolution are likely to be important
in polycrystalline diamond in this regime.
(Fig.~\ref{fig:gpolycmp}.)

\begin{figure}
\begin{center}\includegraphics[scale=0.72]{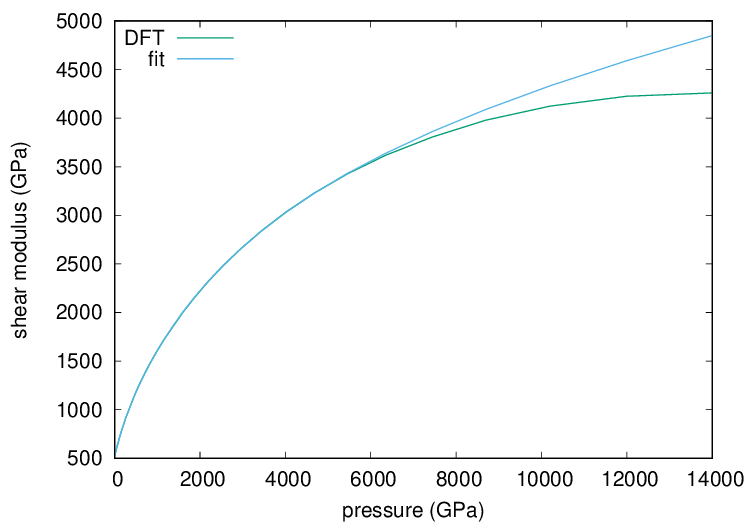}\end{center}
\caption{Shear modulus for equiaxed polycrystalline diamond:
   predicted variation and improved Steinberg-Guinan fit.}
\label{fig:gpolyfit}
\end{figure}

\begin{figure}
\begin{center}\includegraphics[scale=0.72]{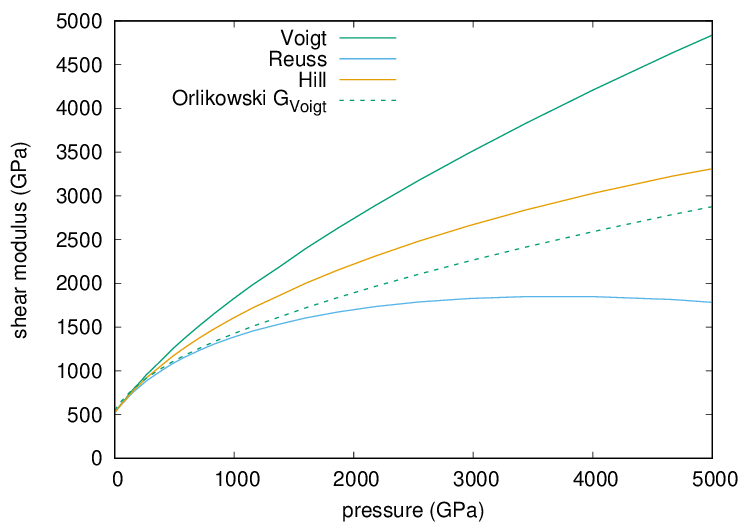}\end{center}
\caption{Predicted shear modulus for equiaxed polycrystalline diamond
   with alternative averaging methods.}
\label{fig:gpolycmp1}
\end{figure}

By fitting the parameters in the improved Steinberg-Guinan model (Table~\ref{tab:isgparams}),
it reproduced the Hill average shear modulus
closely over the full range considered.
The transition function was again broad enough that
$G_h$ affected the low pressure region.
Interestingly, the previous polycrystalline model, which was described as
the Voigt average \cite{Orlikowski2007},
lies close to the present Reuss average to the $\sim$1.5\,TPa 
covered by the previous study.
(Figs~\ref{fig:gpolyfit} and \ref{fig:gpolycmp1}.)

\subsection{Flow stress}
We deduced the flow stress $Y$ by comparing the stress-density response of
diamond deduced from ramp loading experiments with the isentrope.
The ramp experiments used polycrystalline samples, so we used the
Hill polycrystalline average shear modulus $G$.
As discussed above, the variation of $G$ with temperature is usually
less for $G(\rho,T)$ than $G(p,T)$, so we took the polycrystalline
average expressed as $G(\rho)$ and ignored the thermal variation.

\begin{figure}
\begin{center}\includegraphics[scale=0.72]{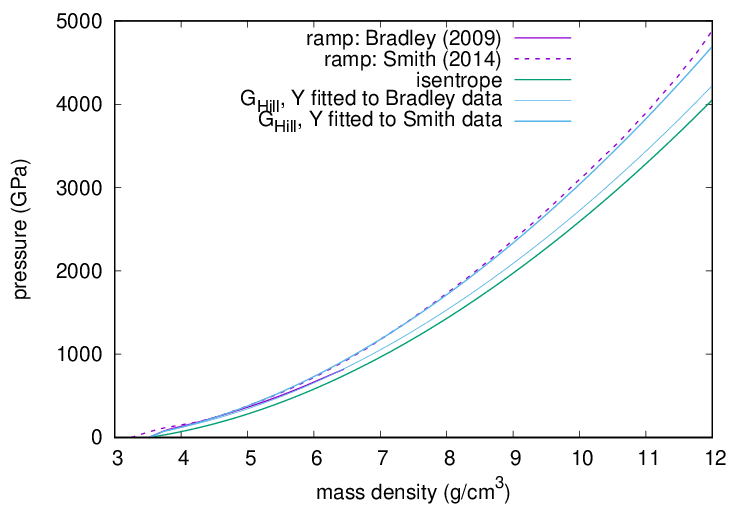}\end{center}
\caption{Comparison of ramp-loading measurements \cite{Bradley2009,Smith2014}
   with simulations assuming different flow stress models.}
\label{fig:strisencmp}
\end{figure}

\begin{figure}
\begin{center}\includegraphics[scale=0.72]{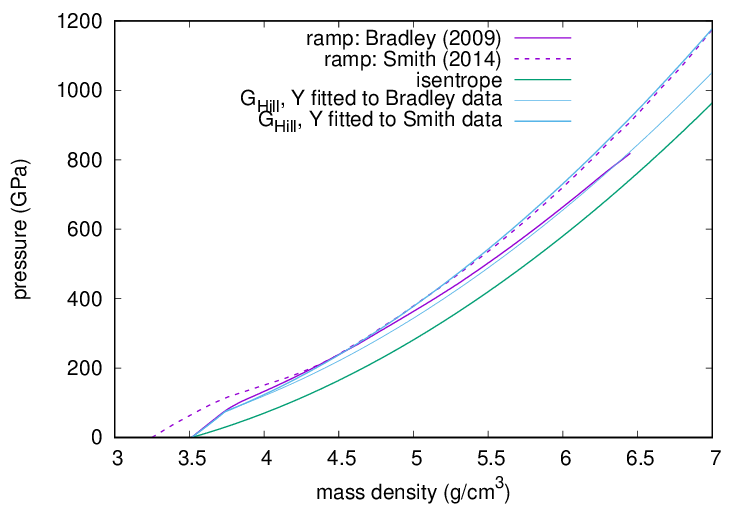}\end{center}
\caption{Comparison of ramp-loading measurements \cite{Bradley2009,Smith2014}
   with simulations assuming different flow stress models
   (detail at low pressure).}
\label{fig:strisencmp1}
\end{figure}

The uniaxial compression measurements from the {\sc omega} and NIF facilities
were taken on samples of different initial density, and the nominal data were not
consistent in the region over which the densities overlapped, although the $1\sigma$
uncertainties overlapped.
The {\sc omega} data were taken on samples of close to full density, which is more
straightforward to interpret,
and ranged up to a normal stress $\sim$0.8\,TPa \cite{Bradley2009}.
The nominal stress-density data were reproduced with a constant flow stress $Y=70$\,GPa.
The $1\sigma$ limits would imply a monotonic increase or decrease of flow stress
with strain.
The NIF data \cite{Smith2014} were taken on nanocrystalline samples relevant for
ICF fuel capsules of significantly lower density, 3.25\,g/cm$^3$.
The nominal stress-density data could be described by a flow stress 
$Y(\rho)=f(\rho) G(\rho)$ where $f$ is a relatively slowly-varying function,
linear from 0.12 at 3.5\,g/cm$^3$, to 0.17 at 10\,g/cm$^3$.
This observation is equivalent to a criterion for plastic flow that there
is a maximum elastic strain that diamond can support which varies slowly
with compression.
Following the lower uncertainty bound in the ramp data, maximum elastic strain
at 10\,g/cm$^3$ would be around 0.07;
following the upper bound, it would be around 0.35, with a variation faster than linear.
The {\sc omega} data are consistent with these relations, within their uncertainty.
(Figs~\ref{fig:strisencmp} and \ref{fig:strisencmp1}.)

\section{Heating on ramp compression}
Diamond is notable for its high yield stress and low heat capacity,
so it is particularly interesting to consider the effect of plastic work
on heating during ramp loading.
The relationship between plastic work and heat generation has not been
thoroughly studied,
and the assumptions made in hydrocode simulations of plastic flow are
rarely even documented.
They vary between assuming that 100\%\ of plastic work appears as heat
to ignoring it altogether.
Some hydrocodes allow a parameter to be set controlling the fraction
of plastic work contributing to heating, the Taylor-Quinney factor, $\fTQ$.
Studies of the heating of bulk metals from plastic working at low rates,
such as occur in manufacturing forming processes, have suggested
$0.7 < \fTQ < 0.95$ \cite{Taylor1934}.
More recent studies have suggested that $\fTQ$ may vary more widely
with material and loading conditions \cite{Zubelewicz2019}.
The underlying process is that plastic deformation occurs through
microstructural changes such as dislocation motion and evolution,
which have an associated change in energy; heating is that part of the
plastic work not accompanied by a change in potential energy of the 
microstructure.

It is known that deformation at high strain rates can induce a very high
dislocation density.
The dislocation population evolves to approach the equilibrium distribution
for the instantaneous strain rate and temperature,
and so dislocations may annihilate over a finite interval after high-rate
deformation, even if the net strain rate for the material is zero.
Thus the Taylor-Quinney factor is not a general representation 
of plastic heating, and $\fTQ$ may even appear to exhibit values outside
the range 0 to 1 \cite{Zubelewicz2019}.
However, a more rigorous representation is not yet available,
particularly for strong covalently-bonded substances such as diamond, and so it is useful
to consider the sensitivity to $\fTQ$ for deformation under simplified
conditions.

\begin{figure}
\begin{center}\includegraphics[scale=0.72]{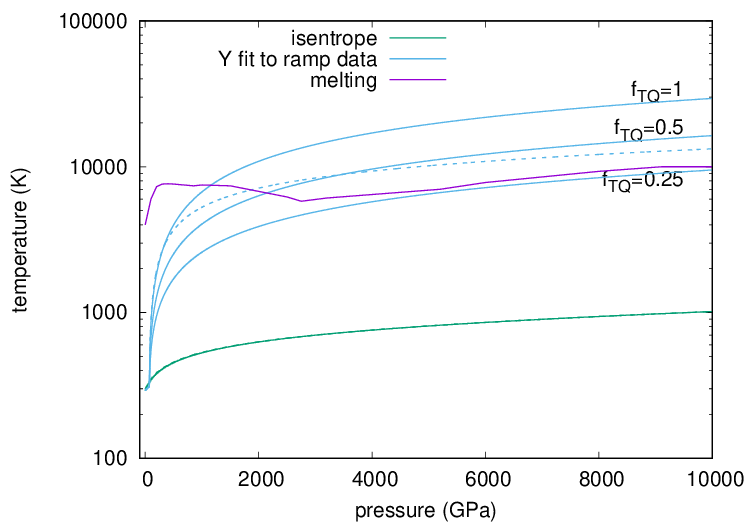}\end{center}
\caption{Predicted temperature during ramp loading, for different
   models of the flow stress: none (green),
   fit to stress measurements of Bradley et al \cite{Bradley2009} (dashed blue),
   and fit to measurements of Smith et al \cite{Smith2014} (solid blue);
   compared with melt locus from multiphase EOS model \cite{Benedict2014}.}
\label{fig:strcmppt}
\end{figure}

In diamond, with the elasticity and flow stress deduced above,
a value $\fTQ=1$ leads to melting on ramp loading to pressures exceeding
1.2\,TPa, using the melt locus from the five-phase EOS model \cite{Benedict2014}
(Fig.~\ref{fig:strcmppt}).
For C to remain solid as inferred from ramp EOS experiments \cite{Smith2014}, 
$\fTQ < 0.25$.
In the five-phase model, melting on this loading path occurs from the BC8 phase.
If diamond persists as a metastable phase,
melting could potentially occur at a lower temperature if kinetically favored
over the transition to BC8;
the elastic strain energy in uniaxially-compressed diamond could lower the
melting point still further.
These contributions to an elevated free energy in dynamically-compressed 
diamond argue for a still-lower value of $\fTQ$.
There may be additional constraints from the observation
that diamond does not transform into other structures to at least 2\,TPa,
as this suggests that the heating is insufficient for the kinetics of
transition to the BC8 structure, with a barrier $\sim$2\,eV,
on $o(1)$\,ns time scales.

\section{Conclusions}
We constructed a DFT-based Mie-Gr\"uneisen EOS model for C in the diamond structure
which is consistent with all high-pressure data and slightly different than
previous EOS models.
We found it necessary to modify the Vinet function used to describe the
cold curve in order to represent the DFT results to high pressure.
The Vashchenko-Zubarev model for the Gr\"uneisen parameter performed 
better than the Dugdale-MacDonald or Slater models.

We also used DFT calculations to predict the variation of the elastic moduli
of diamond.
The predictions were consistent with the previous results of Orlikowski
and Valdez, but extended to much higher pressures.
The calculations of $c_{11}$ and $c_{12}$ were well-behaved, 
and predict an instability to tetragonal distortions above 20\,g/cm$^3$ 
or 12.5\,TPa.
Calculations of $c_{44}$ were noisy above 9\,g/cm$^3$ or 2\,TPa,
which may suggest the onset of a shear distortion.

Effective isotropic
shear moduli were deduced as a function of compression or pressure 
for $\{100\}$, $\{110\}$, and $\{111\}$ crystals. 
The predicted behavior was different than that of Orlikowski's
polycrystalline model, which has been used commonly for simulations of
single-crystal diamond in HED experiments.
The polycrystalline shear modulus was calculated from the single-crystal
moduli.
The Voigt, Reuss, and Hill averages matched at low pressure, 
but deviated significantly at pressures above a hundred gigapascals,
by $\sim$10\%\ around 0.5\,TPa, and reaching 100\%\ by 2.5\,TPa. 
Thus the texture and its evolution are likely to be important 
in polycrystalline diamond.

Analytic models were developed for the elastic and shear moduli 
as a function of mass density or pressure. 
The `improved Steinberg-Guinan' model of shear modulus was found to
represent the Hill average polycrystalline shear modulus
well over the full range considered.

Given the EOS and shear modulus, the flow stress was deduced from
stress-density data obtained in ramp-loading experiments.
Its behavior was consistent with a roughly constant maximum elastic strain
before the onset of flow,
although the uncertainty in the ramp experiments translated to a significant
uncertainty in maximum elastic strain.
The large flow stress implies a significant amount of heating from
plastic work in ramp loading.
If all the plastic work appeared as heat, ramp-loaded samples would melt
below 1.2\,TPa.
The observation of solid diffraction at higher pressures may therefore
be a novel constraint on the Taylor-Quinney factor.

\section*{Acknowledgments}
Dr Florent Occelli kindly provided his original diamond anvil cell data.

This work was performed under the auspices of
the U.S. Department of Energy under contract DE-AC52-07NA27344.

\end{document}